\documentstyle[aps,prl,graphicx]{revtex}

\input psfig.sty
\draft
\begin{document}
\twocolumn 
\wideabs{  
\title{Driving Bose-Einstein condensate vorticity with a rotating normal cloud}
\author{P.~C. Haljan, I. Coddington, P. Engels, and E.~A. Cornell\cite{qpdNIST}}
\address{JILA, National Institute of Standards and Technology and Department of Physics, \\
University of Colorado, Boulder, Colorado 80309-0440}
\date{June 18, 2001}

\maketitle

\begin{abstract}
We have developed an evaporative cooling technique that
accelerates the circulation of an ultra-cold $^{87}$Rb gas,
confined in a static harmonic potential. As a normal gas is
evaporatively spun up and cooled below quantum degeneracy, it is
found to nucleate vorticity in a Bose-Einstein condensate.
Measurements of the condensate's aspect ratio and surface-wave
excitations are consistent with effective rigid-body rotation.
Rotation rates of up to 94$\%$ of the centrifugal limit are
inferred. A threshold in the normal cloud's rotation is observed
for the intrinsic nucleation of the first vortex. The threshold
value lies below the prediction for a nucleation mechanism
involving the excitation of surface-waves of the ground-state
condensate.
\end{abstract}

 } 

\par
To paraphrase an ancient riddle, what happens when an
irresistible torque meets an irrotational fluid? The answer has
been known for more than 50 years: a quantized vortex is
nucleated.  Vortices alone contribute to a superfluid's
rotation, so that the bulk of the fluid may remain curl-free.
The nucleation of vortices in bulk superfluid Helium has been
the topic of extensive study (for a review see
\cite{Donnellyreview}). In the archetypical experiment, a
rotatable pot filled with a mixture of superfluid and normal
liquid Helium undergoes gradual angular acceleration. The normal
fluid and the walls of the pot rotate together as a rigid body,
defining a rotating environment. At some threshold angular
velocity, a vortex line is nucleated at the circumference of the
pot, and then quickly migrates inward until it is collinear with
the axis of rotation. Further angular acceleration results in
the nucleation of more vortices; eventually the fluid is filled
with an array of vortex lines \cite{vortexarray}. A central
theme
\cite{Ruutu1997} of this research is the question: to what
extent is the nucleation process ``extrinsic," {\it i.e.}
dependent on such details as the roughness of the surface of the
walls, and to what extent is it ``intrinsic"
\cite{intrinsicnucleation,intrinsicexperiments}, {\it i.e.} driven (in the limit of
microscopically smooth walls) by the flow of normal fluid along
the boundary of the superfluid? In the analogous
rotating-potential experiments with a dilute-gas Bose-Einstein
condensate (BEC), the confining potential and the normal fluid
typically rotate at different rates
\cite{Guery-Odelin2000a}. In this context, the
extrinsic-intrinsic question can be restated as: is it the
confining potential or the normal fluid that defines the
rotating environment?

\par
Vortices in a BEC have been created with wavefunction
engineering \cite{firstJILAvortexpapers}, through the decay of
solitons \cite{Anderson2001a,HauScience}, and in the wake of
moving objects \cite{Inouyecondmat,recentMITpreprint}. The first
rotating-potential experiment to detect vortices in a BEC was
performed by the Paris group \cite{Madison2000a}; results have
also been obtained by the MIT \cite{Abo-Shaeer2001a} and Oxford
\cite{recentOxfordpreprint} groups. In these experiments
the role of the normal fluid was secondary to that of the
rotating potential; it is conceivable the normal fluid was not
rotating at all. This paper presents vortex nucleation
experiments performed in the opposite limit, namely in the
environment of a rotating normal gas in a static confining
potential. Such an environment allows for the isolated study of
the intrinsic mechanism for vortex nucleation.

\par
Our experiments begin with a magnetically trapped cloud of about
$6\cdot10^{6}$ $^{87}$Rb atoms, in the $|F=1,m_{F}=-1\rangle$
hyperfine state, cooled close to the critical temperature
$T_{c}=67$ nK. The atoms are initially confined in an axially
symmetric, oblate and harmonic potential \cite{Petrich1995a}
with axis of symmetry along the vertical (`z') axis. To induce
rotation of the cloud, we first gradually apply an elliptical
deformation to the potential in its horizontal plane of
symmetry, and then rotate the deformation \cite{Jin1996a} about
the vertical axis at a fixed angular frequency. The rotating
potential is characterized by an axial frequency
$\omega_{z}=2\pi(13.6)$Hz, time-averaged radial frequency
$\langle\omega_{\rho}\rangle=2\pi(6.8)$Hz and a horizontal
ellipticity of $25\%$. Such a large rotating trap asymmetry,
accessible in the oblate configuration of our apparatus, is
found to be necessary not only to get the cloud rotating, but
also to sustain ongoing rotation. Moreover, in steady state the
cloud does not reach the rotation rate of the applied asymmetry.
We believe that the stirring process is fighting a small, static
asymmetry that acts to despin the cloud
\cite{spinupanddowntime}.

\par
In thermal equilibrium, a normal cloud rotates as a rigid body
with the centrifugal force causing the cloud to bulge outwards
in the radial direction. In order to detect the rotation, we use
a nondestructive phase-contrast technique to image the cloud
\textit{in situ} from the side. Four sequential pictures of a
given cloud are taken to average over oscillations in the widths
and to improve the signal to noise. The cloud temperature is
extracted from its vertical width $\sigma_{z}$, which is
unaffected by rotation about the vertical axis. The rotation of
the cloud $\Omega_{N}$ is determined from the aspect ratio,
$\lambda=\sigma_{z}/\sigma_{\rho}$, using the relation:
\begin{equation}
\Omega/\omega_{\rho}=\sqrt{1-(\lambda/\lambda_{o})^{2}}
\label{artorotation}
\end{equation}
where $\lambda_{o}$ is the static aspect ratio. The technique of
side-view imaging is crucial for distinguishing between changes
in radial size due to temperature and to rotation.

\par
\begin{figure}  
\begin{center}
\psfig{figure=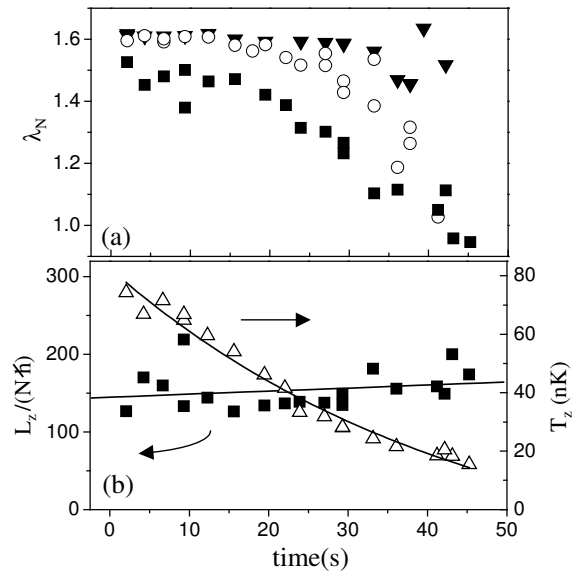,width=1\linewidth,clip=}
\end{center}
\caption {
(a)Aspect ratio of a rotating normal cloud during evaporation
preferentially along the axis of rotation. Three different
initial rotations of the cloud are shown, obtained by first
stirring the cloud for 15s with an applied rotation rate of
0$\omega_{\rho}$ (inverted triangles), 0.07$\omega_{\rho}$ (open
circles) and 0.37$\omega_{\rho}$ (squares). (b)Angular momentum
per particle (squares) and temperature (triangles) of the normal
cloud during evaporation for the 0.37$\omega_{\rho}$ case in
(a).}
\label{1Devaporation}
\end{figure}

For stirring rates up to 2.5 Hz, the rotation of the cloud
reaches its steady-state value by 15 s or less. After 15 s, the
rotating trap asymmetry is ramped off, leaving non-condensed
clouds a factor of 1.2~--~1.3 above $T_{c}$, for stirring
frequencies 0~--~2.5 Hz. Radio frequency (rf) evaporation is
then used to cool the normal cloud to BEC. In the oblate
configuration of our TOP trap \cite{Petrich1995a}, we have found
that rf evaporation immediately quenches the rotation,
presumably because the selection process, which removes atoms
with large radial displacements, preferentially removes atoms
with large axial components of their angular momentum. By
adiabatically distorting the trap into a prolate geometry with
$\{\omega_{\rho},\omega_{z}\}=2\pi\{8.35,5.45\}$Hz
\cite{Ensher1998a}, we can instead cool the cloud by removing
atoms with large {\it axial} displacements and thereby reduce
the effect of the evaporation on the axial angular momentum. As
the normal cloud is evaporated, its aspect ratio is observed to
decrease continuously, indicating a monotonically increasing
rotation rate (Fig.~\ref{1Devaporation}a). During the
evaporation, the angular momentum per particle of the normal
cloud remains roughly constant, even though the number of atoms
is reduced by over a factor of five and temperature, by a factor
of four (Fig.~\ref{1Devaporation}b). As the cloud cools and
shrinks, it must spin up for the angular momentum per particle
to remain fixed.

\par
To reach significant rotation rates by the end of evaporation
requires the lifetime of the normal cloud's angular momentum to
be comparable to the evaporation time. The nearly
one-dimensional nature of the evaporation together with the low
average trap frequencies make cooling to BEC in the prolate trap
very slow ($\sim$50 s). We obtain angular momentum lifetimes
this long by shimming the azimuthal trap symmetry to better than
0.1$\%$.

\par
Towards the end of the evaporation, a condensate begins to
appear at the center of the rotating normal cloud
\cite{nucleationdistinction}. We discuss first the results of
experiments in which we continue the evaporation until little or
no normal fraction remains. In this case, we find that the
rotating normal component has given birth to a condensate
distended in its radial dimension, as one would expect for a
classical rotating body under the influence of the centrifugal
force. This effect is reminiscent of liquid Helium experiments,
in which the surface of a rotating bucket of superfluid exhibits
the same meniscus curvature as for an ordinary viscous fluid
\cite{He4meniscus}. For large enough numbers of vortices in the
condensate, the correspondence principle would suggest that the
rotation field, coarse grained over the cloud, should go over to
the classical limit of rigid-body rotation. In this limit, the
classical Eq.\ (\ref{artorotation}) should connect condensate
aspect ratio to rotation rate.

\par
Alternatively, we can study the angular momentum in the BEC
directly by exciting quadrupolar surface waves
\cite{Chevy2000a,Haljan2001a} with a rotating weak deformity of
the magnetic

\begin{figure}  
\begin{center}
\psfig{figure=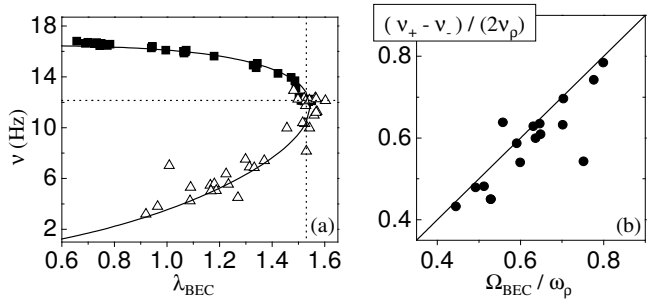,width=1\linewidth,clip=}
\end{center}
\caption {
Quadrupolar surface-wave spectroscopy of condensates formed in a
rotating normal cloud. (a)Quadrupolar frequency as a function of
condensate aspect ratio for the $m=+2$ (squares) and $m=-2$
(triangles) surface waves. Solid lines are a single fit to the
combined data using the theory of Zambelli and Stringari
\protect\cite{Zambelli1998a,Stringarirecent}. Dotted lines indicate
average frequency and aspect ratio for a static BEC. (b)The
splitting between $m=\pm2$ frequencies scaled by twice the trap
frequency, plotted explicitly as a function of BEC rotation rate
inferred from the aspect ratio. The solid line is the prediction
from the same theory as in (a). Each plotted point is obtained
from a single $m=-2$ measurement in (a) combined with a spline
interpolation to the relatively quiet $m=+2$ data. Only aspect
ratios smaller than 1.43 (corresponding to
$\Omega_{BEC}/\omega_{\rho}\geq0.35$) are included to avoid
obtaining imaginary rotation frequencies due to experimental
noise in the aspect ratio.}
\label{quadspectroscopy}
\end{figure}

\noindent
trap. The quadrupolar surfaces waves are characterized by
angular momentum quantum number $m_{z}=\pm2$ where the
$m_{z}=+(-)2$ excitation is taken to be co-(counter-)propagating
with the rotation of the condensate. By varying the initial stir
rate applied to the normal cloud, condensates of different
aspect ratio can be accessed. In Fig.~\ref{quadspectroscopy}a,
the frequency of the $m_{z}=\pm2$ modes is shown as a function
of condensate aspect ratio. The $m_{z}=+2$ mode is seen to speed
up and the $m_{z}=-2$ to slow down due to the presence of
vorticity in the condensate.

\par
For small rotation rates, the splitting between the $m_{z}=\pm2$
modes is predicted to be linearly proportional to the mean
angular momentum of the condensate
\cite{quadtheorypapers,Zambelli1998a}. In the large-$\Omega$
limit of rigid-body rotation, Zambelli and Stringari
\cite{Zambelli1998a,Stringarirecent} have used a sum-rule
argument to show that the splitting between the modes is simply
$2 \Omega$, and, further, that the sum of the squared
frequencies of the two modes is independent of rotation rate. A
best fit of this model to the combined $m=\pm2$ data is shown in
Fig.~\ref{quadspectroscopy}a, where the rigid-body rotation rate
has been inferred from the condensate aspect ratio and the
classical Eq.\ (\ref{artorotation}). Perhaps more intuitively,
the frequency splitting is plotted explicitly versus inferred
rotation rate in Fig.~\ref{quadspectroscopy}b. The excellent
agreement with the model of Zambelli and Stringari is compelling
evidence in favor of the reasonableness of using Eq.\
(\ref{artorotation}) to connect the condensate aspect ratio with
its effective rotation rate. This

\noindent
is further born out by extensive 3-D numerical simulations of
the Gross-Pitaevskii equation for the parameters of our
experiment, by Feder and Clark. Their numerical simulations
confirm that a condensate in an environment rotating at
frequency $\Omega>0.5~\omega_{\rho}$ will equilibrate close to
the aspect ratio given by Eq.~(1)
\cite{Federrecent}.

\par
In pure condensate samples we have observed aspect ratios as
pronounced as 0.35~$\lambda_{o}$, corresponding to a rotation
rate of $0.94~\omega_{\rho}$. The rapid rotation rate, combined
with the increased condensate area arising from its radial
bulge, mean that the condensate must be supporting a large
number of vortices. Feder and Clark \cite{Federrecent} calculate
56. With these initial conditions, we have observed continued
rotation for at least 140 s.

\par
\begin{figure}  
\begin{center}
\psfig{figure=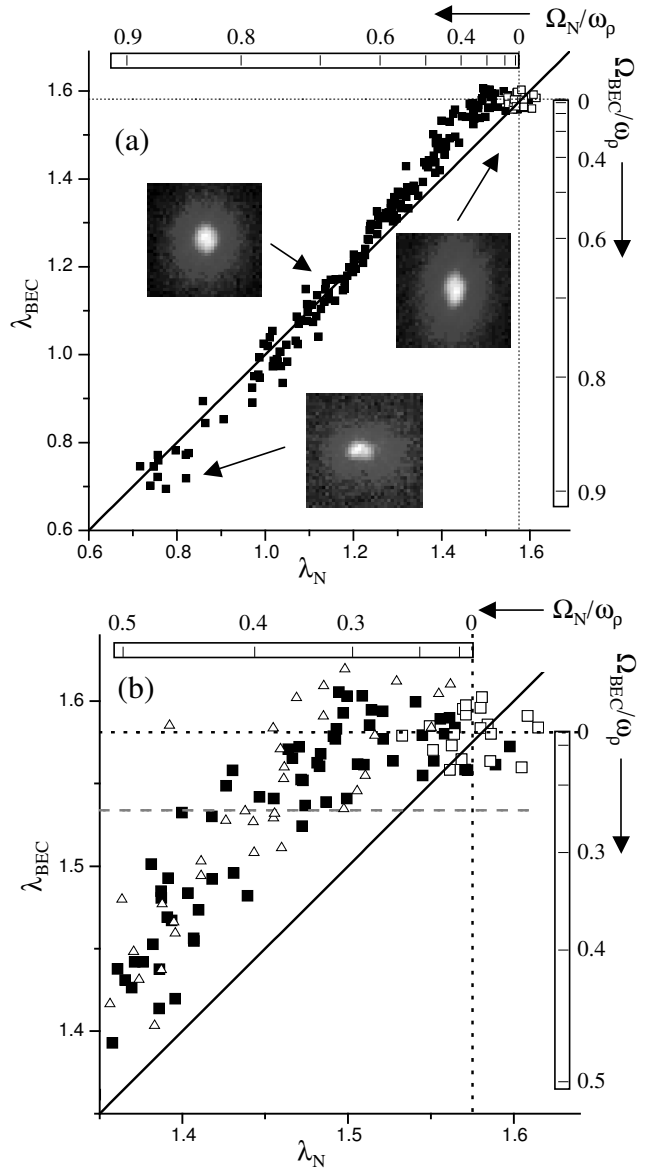,width=1\linewidth,clip=}
\end{center}
\caption {
(a)Aspect ratio of the BEC vs. that of the normal cloud after
evaporation halted. Right and top scales provide a conversion
from aspect ratio to classical rigid-body rotation rate for BEC
and normal cloud respectively. Data for evaporation of both a
static (empty squares) and rotating cloud (filled squares) are
shown. Dotted lines indicate average static aspect ratios for
both BEC and normal cloud. A solid 1:1 line is superimposed on
the data. Three representative integrated density profiles of
two-component clouds indicate the range of different aspect
ratios observed. (b)A magnified version of the region of high
aspect ratio (low rotation rate) in (a). Added to the plot are
data (triangles) obtained with evaporative spin-up 3 times
slower than for the filled squares. A dashed line indicates the
aspect ratio expected for a BEC with a single, centered vortex
as calculated by Feder and Clark
\protect\cite{Federrecent}.}\label{rigidbody}
\end{figure}

If the evaporation is stopped before the normal cloud has been
completely removed, a comparison can be made between the aspect
ratios, and hence rotation rates, of the condensate and normal
cloud. By adjusting the initial stir rate applied to the normal
cloud and the depth of the evaporation, we are able to reach
different rotation rates for a given condensate fraction. After
the evaporation is stopped, a time of 5 s is allowed for the gas
to rethermalize to the last evaporative cut. We then take four
nondestructive pictures and fit the images to a two-component
distribution. Figure \ref{rigidbody} shows a plot of the
condensate aspect ratio $\lambda_{BEC}$ compared with the aspect
ratio $\lambda_{N}$ of the normal cloud where each point
represents a single realization of the experiment.

\newpage
\par
For $\lambda_{N}<1.36$ ($\Omega_{N}>0.5~\omega_{\rho}$), the
condensate aspect ratio closely tracks the normal aspect ratio,
providing a further manifestation of the correspondence
principle for a highly rotating BEC. In the vicinity of
$\lambda_{N}=1.48$ ($\Omega_{N}=0.35~\omega_{\rho}$), we observe
threshold behavior in the condensate rotation. At this low value
of rotation we don't expect the rigid-body model to be valid for
the condensate; but we make use of a numerical calculation by
Feder and Clark \cite{Federrecent} to indicate the change in
condensate aspect ratio associated with the presence of a
single, centered vortex (Fig.~\ref{rigidbody}b). There is
considerable scatter in the data so one cannot make a strong
statement about the nature of the threshold shape, but clearly,
somewhere between $0.32<\Omega_{N}/\omega_{\rho}<0.38$ the first
vortex is nucleated.

\par
A comparison of this threshold location with two theoretical
rotation rates for our particular cloud provides some insight
into the nature of the vortex nucleation. The first value is the
threshold for the thermodynamic stability of a single vortex.
For our experiment this number is 0.2~--~0.25 $\omega_{\rho}$
\cite{thermotheory}, distinctly lower than our observed
threshold for nucleation. The second value is $\omega_{min}$,
the frequency at which the slowest surface-wave mode propagates
around the circumference of the condensate. The Paris group has
shown that for their "extrinsic" nucleation process (vortices
nucleated by a rotating asymmetric potential) the key mechanism
is the nonlinear excitation of surface waves
\cite{Madison2000a,Madison2001}. Results from MIT
\cite{recentMITpreprint,Abo-Shaeer2001a} and from Oxford
\cite{recentOxfordpreprint} are also consistent with such a
mechanism. For the parameters of our experiment,
$\omega_{min}=0.4~\omega_{\rho}$
\cite{Dalfovo2000b}. Our observed threshold for ``intrinsic"
nucleation is clearly under this value, and thus we cannot
interpret our effect in terms of a normal ``wind" exciting
surface waves on our condensate.

\par
The data presented in Fig.~\ref{rigidbody} include a range of
condensate fraction from 0.1-0.6 for each rotation rate of the
normal cloud, although no segregation relative to either axis is
evident for plots of different BEC fractions. Moreover, by
reducing the rate of evaporation, we have decreased the rate of
acceleration of the normal cloud rotation by a factor of 3, and
still observe a threshold for vortex formation between 0.32 and
0.38 (Fig.~\ref{rigidbody}b).

\par
In future work we plan to use the controlled rotation of the
normal cloud to study the life cycle of a BEC's vorticity
\cite{Fedichev2001a}, including any hysteresis between spin-up
and spin-down due to vortex metastability. By halting the
rotation of the normal cloud, we have already found that we can
reduce the lifetime of the BEC's rotation by a factor of 10.
Finally, the rotating normal cloud can create equilibrated
condensates with very large rotation rates, which may allow us
to approach the regime for which the vortices are so
close-packed that their separation becomes comparable to the
healing length
\cite{Ho2001a}.

\par
We acknowledge very useful conversations with
Sea{\nolinebreak}mus Davis, David Feder, Sandro Stringari, and
Carl Wie{\nolinebreak}man. This work was supported by NSF and
NIST funding. P.~E. acknowledges support from the
Alexander-von-Humboldt foundation.


%


\begin{thebibliography}{10}
\bibitem{nucleationdistinction}
In L.He. experiments, one expects very different behavior if one
either (i)
  cools below $T_{c}$ before crossing the threshold rotation for vortex
  nucleation or (ii) vice versa. In our experiment, the harmonic
  ``pot{\protect"} means that case (i) applies: the new-born condensate is
  initially too small to support a vortex.
\end{thebibliography}
\end{document}